%% file: main.tex
\documentclass[epj,nopacs]{svjour}

\usepackage[T1]{fontenc}
\usepackage[utf8]{inputenc}
\usepackage{microtype}
\usepackage[htt]{hyphenat}
\usepackage{amsmath,amssymb}
\usepackage{graphicx}
\usepackage[usenames,dvipsnames]{xcolor}
\usepackage{color}
\usepackage{orcidlink}
\usepackage{caption}
\usepackage{subcaption}
\usepackage{cite}
\usepackage{hyperref}
\usepackage[normalem]{ulem}
\usepackage{physics}
\usepackage{placeins}
\usepackage{comment}
\usepackage{booktabs, multirow}
\usepackage{enumitem}
\usepackage{bm}
\usepackage{bigints}
\usepackage[nolist,nohyperlinks]{acronym}

\hypersetup{
  colorlinks   = true,
  urlcolor     = blue,
  linkcolor    = blue,
  citecolor   = blue
}
\allowdisplaybreaks[1]
\usepackage{float}

\journalname{Eur. Phys. J. A}

\newcommand{\potsdam}{Institut f{\"u}r Physik und Astronomie, Universit{\"a}t Potsdam, Haus 28, Karl-Liebknecht-Str. 24/25, 14476, Potsdam, Germany}
\newcommand{\aei}{Max Planck Institute for Gravitational Physics (Albert Einstein Institute), Am M{\"u}hlenberg 1, Potsdam 14476, Germany}
\newcommand{\grasp}{Institute for Gravitational and Subatomic Physics (GRASP), Utrecht University, Princetonplein 1, 3584 CC Utrecht, The Netherlands}
\newcommand{\nikhef}{Nikhef, Science Park 105, 1098 XG Amsterdam, The Netherlands}

\newcommand{\Msol}{\rm{M}_\odot}
\newcommand{\Msun}{\Msol}
\newcommand{\MTOV}{M_{\rm{TOV}}}
\newcommand{\nsat}{n_{\rm{sat}}}

\begin{document}

\title{Analyzing GW231109\_235456 and understanding its potential implications for population studies, nuclear physics, and multi-messenger astronomy}

\author{
Thibeau Wouters~\orcidlink{0009-0006-2797-3808}\inst{1,2}
\and
Anna Puecher~\orcidlink{0000-0003-1357-4348}\inst{3}
\and
Peter T. H. Pang~\orcidlink{0000-0001-7041-3239}\inst{2,1}
\and
Tim Dietrich~\orcidlink{0000-0003-2374-307X}\inst{3,4}
}

\institute{
\grasp \and
\nikhef \and
\potsdam \and
\aei
}

\date{Received: date / Revised version: date}

\titlerunning{Analyzing GW231109\_235456}
\authorrunning{T. Wouters et al.}

\abstract{
We study the gravitational-wave trigger GW231109\_235456, a sub-threshold binary neutron star merger candidate observed in the first part of the fourth observing run of the LIGO-Virgo-KAGRA collaboration.
Assuming the trigger is of astrophysical origin, we analyze it using state-of-the-art waveform models and investigate the robustness of the inferred source parameters under different prior choices in Bayesian inference.
We assess the implications for population studies, nuclear physics, and multi-messenger astronomy.
Analyzing the component masses, we find that GW231109\_235456, if astrophysical, supports the proposed double Gaussian mass distribution of neutron stars. Moreover, we find that the remnant most likely collapsed promptly to a black hole and that, because of the large distance, a possible kilonova connected to the merger was noticeably dimmer than AT2017gfo but within reach of observatories such as LSST and Roman.
In addition, we provide constraints on the equation of state from the candidate GW231109\_235456 alone, as well as combined with GW170817 and GW190425.
In our projections for the future, we simulate a similar event using the planned next generation of ground-based gravitational-wave detectors. Our findings indicate that we can constrain the tidal deformability of a $1.4 \ \Msun$ neutron star to about $10\%$, depending on the specific designs and network considered.
}

\maketitle

\section{Introduction}

The first multi-messenger observation of a \ac{BNS} merger, the joint observation of the \ac{GW} signal GW170817~\cite{LIGOScientific:2017vwq}, the kilonova AT2017gfo~\cite{LIGOScientific:2017pwl,Andreoni:2017ppd,Coulter:2017wya,Cowperthwaite:2017dyu,Drout:2017ijr,Lipunov:2017dwd,Shappee:2017zly,Tanvir:2017pws,Troja:2017nqp,J-GEM:2017tyx,Villar:2017wcc}, the gamma-ray burst GRB170817A~\cite{LIGOScientific:2017zic,Goldstein:2017mmi,Savchenko:2017ffs} and its afterglow~\cite{Hallinan:2017woc,Alexander:2018dcl,Margutti:2018xqd,Ghirlanda:2018uyx,Troja:2017nqp,DAvanzo:2018zyz}, showcased the importance of such observations for our understanding of cosmology and the nuclear equation of state~\cite{LIGOScientific:2017adf,LIGOScientific:2018cki}.

Since then, there has not been a definitive multi-messenger detection of a \ac{BNS} merger.
However, a variety of GW and \ac{EM} signals have been observed that are likely connected to \ac{BNS} mergers.
Among them, the GW event GW190425~\cite{LIGOScientific:2020aai}, a compact binary merger with a total mass of $\sim3.4 \Msun$, has been identified as a \ac{BNS} candidate, although the \ac{NSBH} hypothesis cannot be ruled out~\cite{Foley:2020kus,Han:2020qmn,Kyutoku:2020xka,Dudi:2021abi}.
On the \ac{EM} side, GRB211211A~\cite{Rastinejad:2022zbg,Troja:2022yya,Kunert:2023vqd} and GRB230307A~\cite{JWST:2023jqa} have been identified as long GRBs with strong evidence of kilonova components.
However, due to the larger distance of the event and the fact that Advanced LIGO~\cite{LIGOScientific:2014pky} and Advanced Virgo~\cite{VIRGO:2014yos} were not operational at the time, it cannot be definitively confirmed that these events originated from a compact binary merger.

Recently, the fourth Gravitational-Wave Transient Catalog (GWTC-4.0)~\cite{LIGOScientific:2025slb} reported observations from the first part of the fourth observing run (O4a) of the LIGO-Virgo-KAGRA (LVK) collaboration, but did not include any confident \ac{BNS} candidates.
However, a possible sub-threshold candidate, GW231109\_235456 (from here onwards, abbreviated as GW231109), has been identified as a significant trigger in the Hanford and Livingston detectors in a sub-threshold targeted search~\cite{Niu:2025nha}.
This candidate was also reported during online searches on \textsc{GraceDB}~\cite{gracedb} as S231109ci, with a \ac{FAR} of $13.5$ yr${}^{-1}$ and $p_{\rm{astro}} = 0.04$, implying a dominant probability of the signal being of terrestrial origin, i.e., $p_{\rm{terr}} = 0.96$~\cite{Farr:2013yna,LIGOScientific:2016kwr,LIGOScientific:2016ebi}.
In the follow-up search by Ref.~\cite{Niu:2025nha}, a template bank was constructed to target \ac{BNS} systems with \textsc{GstLAL}~\cite{Cannon:2012zt,Messick:2016aqy,Sachdev:2019vvd,Hanna:2019ezx,Cannon:2020qnf,Sakon:2022ibh,Ray:2023nhx,Tsukada:2023edh,Ewing:2023qqe,Joshi:2025zdu,Joshi:2025nty}, using a mass distribution informed by radio observations of double neutron star systems.
As a result, Ref.~\cite{Niu:2025nha} found a nearly $60\%$ improvement of sensitivity to the \ac{BNS} sources targeted by their mass distribution compared to the standard stellar-mass search performed in O4a~\cite{LIGOScientific:2025yae}.
The search reported GW231109 with a \ac{FAR} of $1$ per $10$ years and a \ac{SNR} of $9.7$, and follow-up parameter estimation was performed on the candidate.
While a potential candidate \ac{EM} counterpart was reported by Ref.~\cite{Li:2025rmj}, it is uncertain whether it originates from the same candidate source as GW231109.
Therefore, no consensus has yet been reached about the astrophysical nature of the GW231109 trigger.

Despite the current uncertainty of the astrophysical origin of the trigger, new observations of \ac{BNS} systems are of paramount importance, given their low occurrence rates.
Therefore, in this work, we provide a thorough analysis of the sub-threshold GW candidate GW231109 to determine the properties of the source and examine its consistency with different population models.
We investigate potential constraints on the \ac{EOS} of dense nuclear matter, verifying parameter estimation consistency for different prior choices.
In addition, we estimate the fate of the merger remnant, the properties of the ejecta, and predict the kilonova lightcurves.
Note that these downstream analyses necessarily assume GW231109 to be an astrophysical signal (in particular, a \ac{BNS} merger), an assumption that is, as described above, still subject to debate.
Finally, we simulate a system similar to GW231109 as observed by the third-generation \ac{GW} detectors \ac{ET}~\cite{Punturo:2010zz,Hild:2010id} and \ac{CE}~\cite{Reitze:2019iox,Evans:2021gyd,Evans:2023euw} to project how the increased sensitivity of these detectors would constrain the \ac{EOS} from a similar source.

\section{Gravitational wave parameter estimation}

Using Bayes' theorem, we sample the posterior distribution of the \ac{GW} source parameters with nested sampling~\cite{Skilling:2004pqw,Skilling:2006gxv}.
In particular, we use the \texttt{dynesty} sampler~\cite{Speagle:2019ivv,sergey_koposov_2024_12537467} with $2000$ live points, employing the standard settings of \texttt{bilby}~\cite{Ashton:2018jfp}.
Moreover, we use the most recent version of \texttt{bilby} which corrected a bug in the likelihood function~\cite{Talbot:2025vth}.

Compared to the analysis presented in Ref.~\cite{Niu:2025nha}, we introduce the following changes. First, we adopt more advanced \ac{BNS} waveform models, namely, \texttt{IMRPhenomXAS\_\allowbreak NRTidalv3} for aligned-spin systems and \texttt{IMRPhenomXP\_\allowbreak NRTidalv3} for systems with precessing spins~\cite{Abac:2023ujg,Colleoni:2023ple}.
These models incorporate the \texttt{NRTidalv3} tidal phase calibrated to unequal-mass systems with dynamical tides~\cite{Abac:2023ujg}. Second, we extend the analysis duration to $256$ seconds and increase the maximum frequency to $2048$ Hz to improve sensitivity to tidal effects, which are stronger at higher frequencies~\cite{Dietrich:2020eud,Chatziioannou:2020pqz}.\footnote{The on-source \ac{PSD} is computed with \textsc{BayesWave}~\cite{Cornish:2014kda,Littenberg:2014oda,Cornish:2020dwh} with open-data strain from \textsc{gwosc}~\cite{LIGOScientific:2025snk,KAGRA:2023pio,LIGOScientific:2019lzm}.}
Finally, we accelerate the likelihood evaluation using multibanding~\cite{Garcia-Quiros:2020qlt,Morisaki:2021ngj} rather than \ac{ROQ}~\cite{Smith:2016qas,Morisaki:2023kuq}.

\subsection{Priors}\label{ssec: GW PE priors}

To isolate the effect of priors, we test different prior choices for the masses, spins, and tidal deformabilities of the two \acp{NS}.
For the extrinsic parameters, the priors are identical to Ref.~\cite{Romero-Shaw:2020owr}, with the luminosity distance prior uniform in comoving volume and source frame time.

\textit{Masses:}
We consider four mass prior choices for this analysis.
First, the `default' prior samples uniformly over detector-frame component masses while restricting the detector-frame chirp mass $\mathcal{M}_c$ to the range $[1.29, 1.32]\,\Msun$, and the mass ratio $q = m_2/m_1$ to the range $[0.125, 1]$, following standard \ac{GW} inference practice.\footnote{The specific range for the chirp mass was chosen in order to contain the value of $1.306\,\Msun$ recovered by the template search~\cite{Niu:2025nha}.}
Second, the `uniform' prior adopts a uniform distribution over source-frame component masses in $[1, 3]\,\Msun$, consistent with the \ac{NS} population observed in \acp{GW}~\cite{Landry:2021hvl,KAGRA:2021duu}.
Third, the `Gaussian' population prior~\cite{Ozel:2012ax,Kiziltan:2013oja,Ozel:2016oaf} draws source-frame component masses from a Gaussian distribution $\mathcal{N}\left(1.33\,\Msun, (0.09\,\Msun)^2\right)$.
Fourth, the `double Gaussian' population prior~\cite{Schwab:2010jm,Antoniadis:2016hxz,Alsing:2017bbc,Farrow:2019xnc,Shao:2020bzt,Horvath:2020dkr} draws the source-frame masses from a weighted mixture\footnote{See Ref.~\cite{Shao:2020bzt} for the definition of the relative weight.} of two Gaussian distributions, i.e., $\mathcal{N}(1.372\,\Msun, \left(0.05768\,\Msun\right)^2)$ with a relative weight of $0.7137$, and $\mathcal{N}(1.534\,\Msun, \left(0.09102\,\Msun\right)^2)$ with a relative weight of $0.2863$, which is identical to Ref.~\cite{Niu:2025nha}.

\textit{Spins:}
For the magnitude of the dimensionless component spins $a_i$, we consider two uniform priors where the maximum magnitude is either $0.05$ or $0.4$, referred to as the low-spin and high-spin priors, respectively.
If not specified otherwise, our fiducial runs use the low-spin prior $a_i < 0.05$.
Although the maximum known spin for an \ac{NS} in a binary system is $\sim 0.2$ \cite{Hessels:2006ze}, the limit used in the high-spin prior corresponds to the maximum spin observed in isolated \acp{NS}~\cite{Hessels:2006ze}.
Note that theoretical calculations suggest that dimensionless spins can exceed this value \cite{Dietrich:2015pxa}.

\textit{Tidal deformabilities:}
The information about matter effects is primarily encoded in the dimensionless tidal deformability parameters $\Lambda_i$, which describe the deformation of an \ac{NS} in a binary system as a consequence of the gravitational field created by its companion~\cite{Hinderer:2009ca,Damour:2009vw,Damour:2012yf}.
We consider three priors for the tidal deformabilities in this work.
First, we sample $\Lambda_{i}$ uniformly in the range $[0,5000]$.
Second, we use \ac{QUR}, which are relations between source parameters that are largely independent of the specific \ac{EOS}.
Specifically, we use the binary Love relations~\cite{Yagi:2015pkc} and follow the approach outlined in Ref.~\cite{Chatziioannou:2018vzf}, while acknowledging the associated caveats and drawbacks discussed in Ref.~\cite{Kastaun:2019bxo}.
Finally, we perform parameter estimation by sampling a set of \acp{EOS} constrained by observations.
In particular, we use the set of \acp{EOS} from Ref.~\cite{Koehn:2024set} (with data available at Ref.~\cite{eos_tool}) and the weights derived by the observations collected in `Set A'.\footnote{`Set A' contains information obtained from chiral effective field theory, perturbative quantum chromodynamics, radio measurements of massive pulsars, the X-ray NICER measurement of PSR J0030+0451 and PSR J0740+6620, and the analysis of GW170817; see Ref.~\cite{Koehn:2024set} for details.}
At each likelihood evaluation, the sampled \ac{EOS} is used to compute $\Lambda_{i}$ following the implementation in Ref.~\cite{Pang:2022rzc}.

Our fiducial prior choice for the individual analyses discussed in the following depends on the specific question being addressed.
For inferring the \ac{EOS} in Sec.~\ref{sec:eos}, we use posteriors obtained by sampling $\Lambda_i$ uniformly in the range $[0, 5000]$, to guarantee that the prior remains agnostic to existing \ac{EOS} constraints.
In contrast, for predicting the fate of the remnant (Sec.~\ref{sec:remn}) and potential kilonova lightcurves (Sec.~\ref{ssec: KN LC}), we restrict ourselves to posteriors that perform \ac{EOS} sampling to ensure that the masses and tidal deformabilities samples are consistent with physical constraints imposed by the \ac{EOS}.

\begin{figure}[t]
    \centering
    \includegraphics[width=\columnwidth]{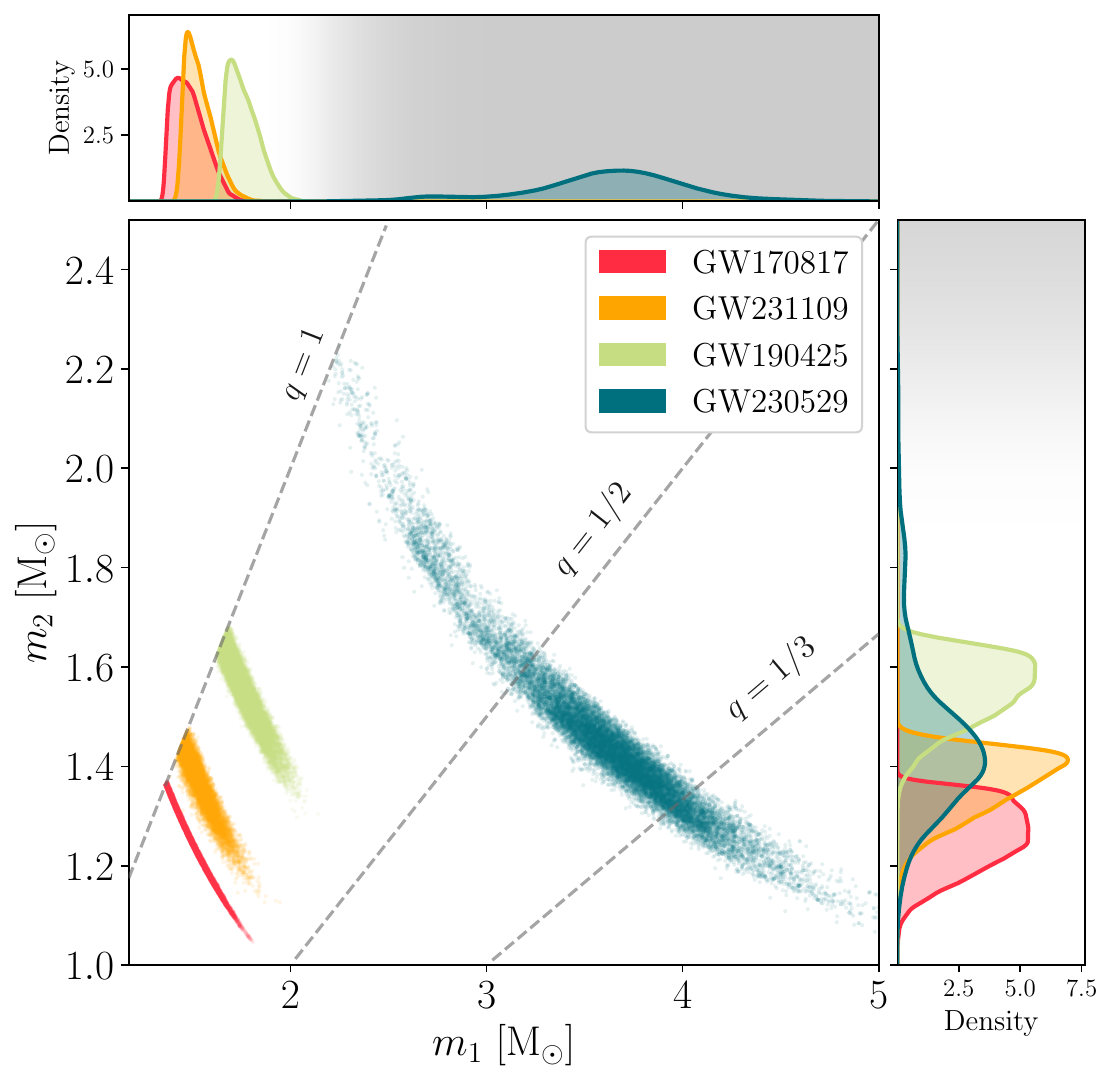}
    \caption{Source-frame component masses of low-mass \ac{GW} events that most likely contain at least one NS, namely, GW170817~\cite{LIGOScientific:2017vwq}, GW190425~\cite{LIGOScientific:2020aai}, GW230529~\cite{LIGOScientific:2024elc}, and GW231109\_235456 (GW231109). GW231109's individual masses lie between those of GW170817 and GW190425.
    The gray shade in the 1D panels shows the transition from NS to BH masses: the color opacity corresponds to the cumulative density function of the TOV mass posterior, based on the uncertainty in the TOV mass inferred later in this work from measurements of heavy pulsars (see Sec.~\ref{sec:eos} for details).}
    \label{fig: m1-m2 overview of BNS events and GW230529}
\end{figure}

\subsection{Source properties}\label{ssec:pe_res}

\begin{figure*}[t]
     \centering
     \begin{subfigure}[b]{0.49\textwidth}
         \centering
         \includegraphics[width=0.99\textwidth]{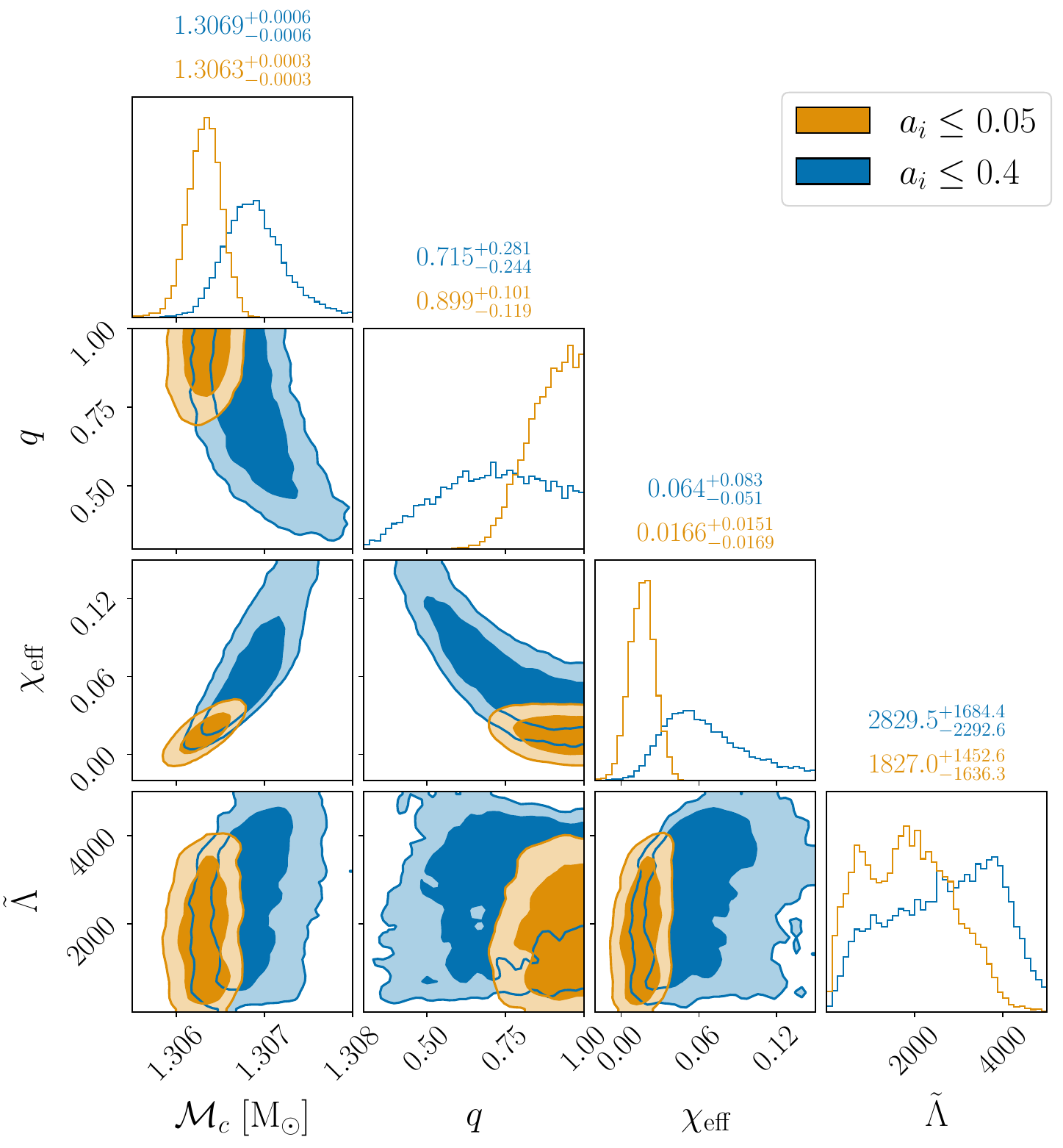}
     \end{subfigure}
     \hfill
     \begin{subfigure}[b]{0.49\textwidth}
         \centering
         \includegraphics[width=0.99\textwidth]{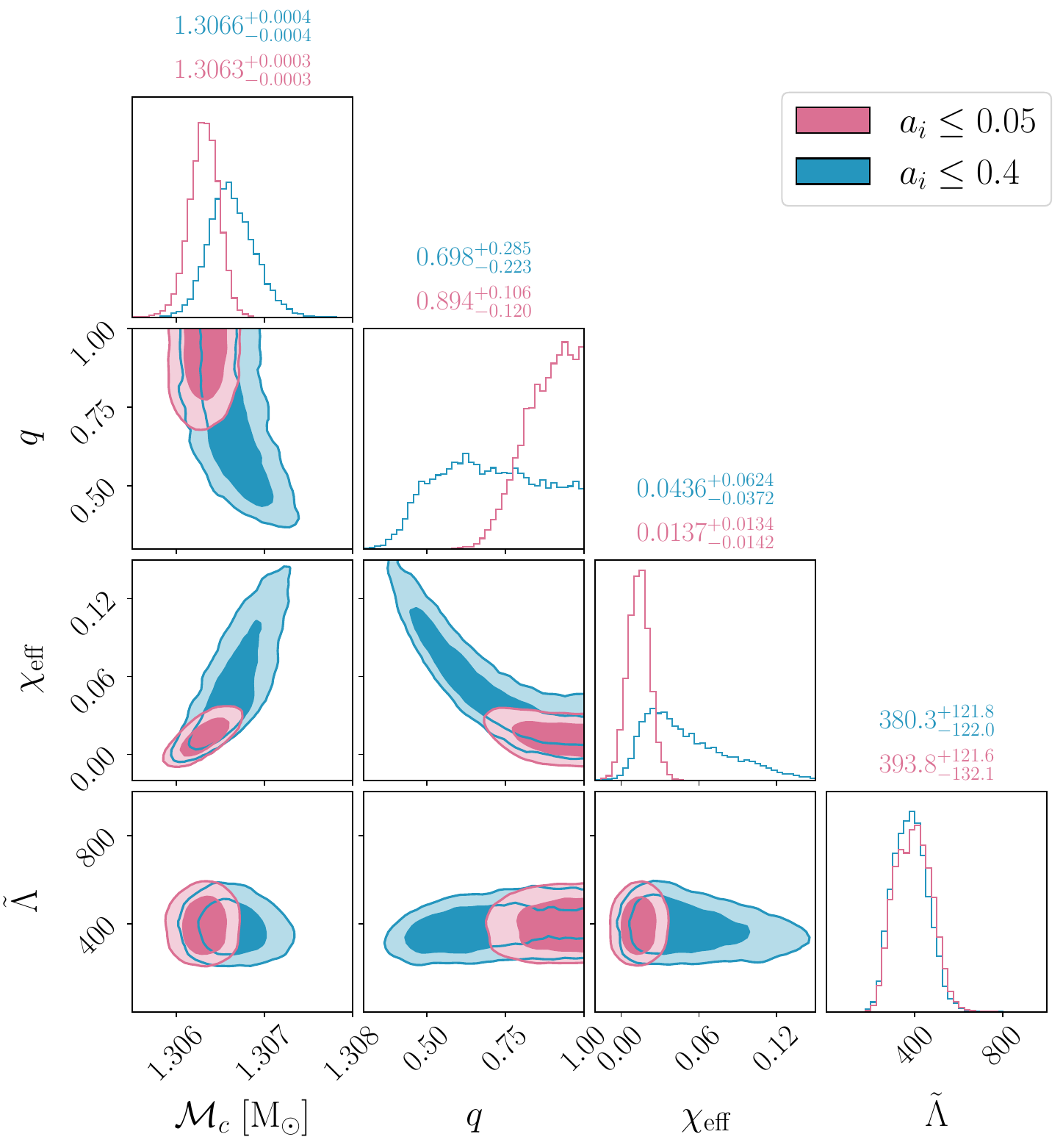}
     \end{subfigure}
        \caption{Posterior on chirp mass $\mathcal{M}_c$, mass ratio $q$, effective spin $\chi_{\rm eff}$, and mass-weighted tidal deformability $\tilde{\Lambda}$ of the \ac{GW} inference using the default mass priors, i.e., uniform in detector-frame component masses. \textit{Left panel}: $\Lambda_i$ are sampled uniformly in the range $[0, 5000]$. \textit{Right panel}: $\Lambda_i$ are determined by the \ac{EOS} sampled on-the-fly, with samples where the primary mass exceeds the TOV mass being discarded. The dark (light) shading indicates the $68\%$ ($95\%$) credible area. The median values of the recovered parameters, together with the $90\%$ credible intervals, are reported above the marginalized posterior distributions.}
        \label{fig: GW PE cornerplots}
\end{figure*}

Figure~\ref{fig: m1-m2 overview of BNS events and GW230529} shows the one- and two-dimensional posteriors for the source-frame component masses recovered for GW231109, compared to other low-mass \ac{GW} observations that likely contained at least one \ac{NS}.
For GW170817, GW190425, and GW231109, the default mass, low-spin, and uniform in $\Lambda_i$ priors were used.
For GW230529, we vary the magnitude of the spin of the \ac{BH} up to $0.99$.
The masses of GW231109 are consistent with those expected for a \ac{BNS} system, being slightly heavier than those of GW170817 but still lighter than those of GW190425.

Figure~\ref{fig: GW PE cornerplots} shows the posteriors for chirp mass $\mathcal{M}_c$, mass ratio $q$, effective spin $\chi_{\rm eff}$~\cite{Santamaria:2010yb,Ajith:2009bn}
\begin{equation}
    \chi_{\rm eff} = \frac{\chi_1 m_1 + \chi_2 m_2}{m_1 + m_2} \, ,
\end{equation}
where $\chi_i$ are the component aligned spins, and mass-weighted tidal deformability~\cite{Flanagan:2007ix,Favata:2013rwa}
\begin{equation}
    \tilde{\Lambda} = \frac{16}{3} \frac{\left(m_1 + 12 m_2\right) m_1^4 \Lambda_1 + \left(m_2 + 12 m_1\right) m_2^4 \Lambda_2}{\left(m_1 + m_2\right)^5} \, .
\end{equation}

The left panel of Fig.~\ref{fig: GW PE cornerplots} shows the posterior with $\Lambda_i$ sampled uniformly in $[0, 5000]$, while the right panel is obtained by sampling the \ac{EOS}; cf.\ Sec.~\ref{ssec: GW PE priors}.
In both cases, we show our different spin prior choices and we use the default mass priors as defined in Sec.~\ref{ssec: GW PE priors}.
We find consistent results across the different mass prior choices.
For the posteriors obtained from sampling the \ac{EOS}, we remove posterior samples for which the primary mass $m_1$ exceeds the \ac{TOV} mass of the corresponding \ac{EOS}.

We find that the broader spin priors result in wider posteriors on the masses, regardless of the prior choice on the tidal deformabilities. This is due to the correlation between mass ratio and spin~\cite{Ng:2018neg}.
When using a larger spin prior, we observe stronger support for high positive spin values.
In these cases, the inspiral process is generally decelerated due to the repulsive spin-orbit interactions that occur at the 1.5 post-Newtonian order.
However, this deceleration can be offset if the system's mass ratio is lower.
Therefore, systems that are both highly spinning and more asymmetric can generate similar \ac{GW} signatures.
Additionally, attractive tidal effects can also help counterbalance the higher spins.
Indeed, the posterior using the uniform prior on tidal deformabilities recovers higher tidal deformabilities when using the high-spin prior.

When directly sampling over the \ac{EOS}, we obtain slightly narrower posteriors on the masses and, consequently, on the spins.
The posterior on $\tilde{\Lambda}$ coincides quite well for both spin priors when sampling the \ac{EOS}, despite the large spin prior resulting in a broader posterior on the mass ratio.
However, the tighter constraints on $\tilde{\Lambda}$ relative to the default prior runs stem from the information already incorporated into the \ac{EOS} prior employed.
Therefore, also in this case, the posterior on $\tilde{\Lambda}$ is dominated by the prior.

Across all priors considered, the signal hypothesis is favoured over the stationary, Gaussian noise hypothesis with log Bayes factors ($\log \mathcal{B}_N^S$)~\cite{Veitch:2008ur,Veitch:2009hd} ranging from $9.4$ to $14.1$.
The lowest $\log \mathcal{B}_N^S$ is achieved with the uniform mass priors (due to their increased prior volumes, which act as Occam penalties), whereas the `default' priors reach $\log \mathcal{B}_N^S = 12.11$.
The highest Bayesian evidence is achieved by the priors that use high spins, the \ac{EOS}-informed prior on tidal deformabilities, or both combined, although the Bayes factors between these prior choices are uninformative.
We note that these Bayes factors should not be taken as decisive evidence for a signal in the data.
Since real detector noise is not necessarily Gaussian or stationary, the noise hypothesis in this model comparison is flawed when applied to real data.
Moreover, the Bayes factor as a detection statistic does not have a calibrated \ac{FAR}, unlike traditional searches.

\begin{figure}[t]
    \centering
    \includegraphics[width=\columnwidth]{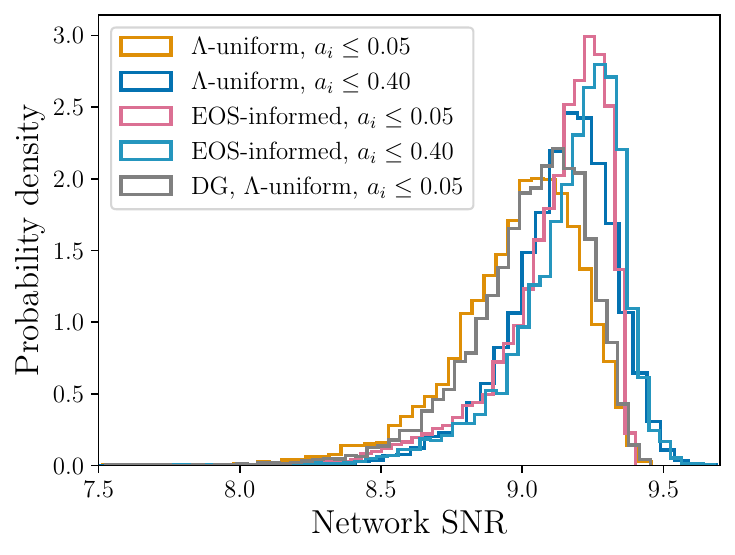}
    \caption{Distribution of network \ac{SNR} for the four prior choices considered in Fig.~\ref{fig: GW PE cornerplots} with default mass prior, and one with the double Gaussian (DG) mass prior.}
    \label{fig: SNR}
\end{figure}

The network \ac{SNR} for different prior choices, shown in Fig.~\ref{fig: SNR}, follows the trends laid out by the Bayes factors.
In this figure, we consider the four analyses presented in Fig.~\ref{fig: GW PE cornerplots} using the default mass priors and various spin and tidal deformability priors, as well as one analysis with the double Gaussian (DG) mass prior, using low-spin and uniform tidal deformability priors.
While the double Gaussian mass prior has a small impact on the network \ac{SNR} distribution, the distributions reach higher medians for priors using high spins, information from the \ac{EOS}, or both.
Moreover, the high-spin priors reach higher maximal values, regardless of the prior choice on the tidal deformability.
This \ac{SNR} inferred from the various prior choices is consistent with other subthreshold triggers reported in GWTC-4.0~\cite{LIGOScientific:2025slb}.

\subsection{Consistency with populations}\label{ssec: populations}

\begin{figure}[t]
    \centering
    \includegraphics[width=\columnwidth]{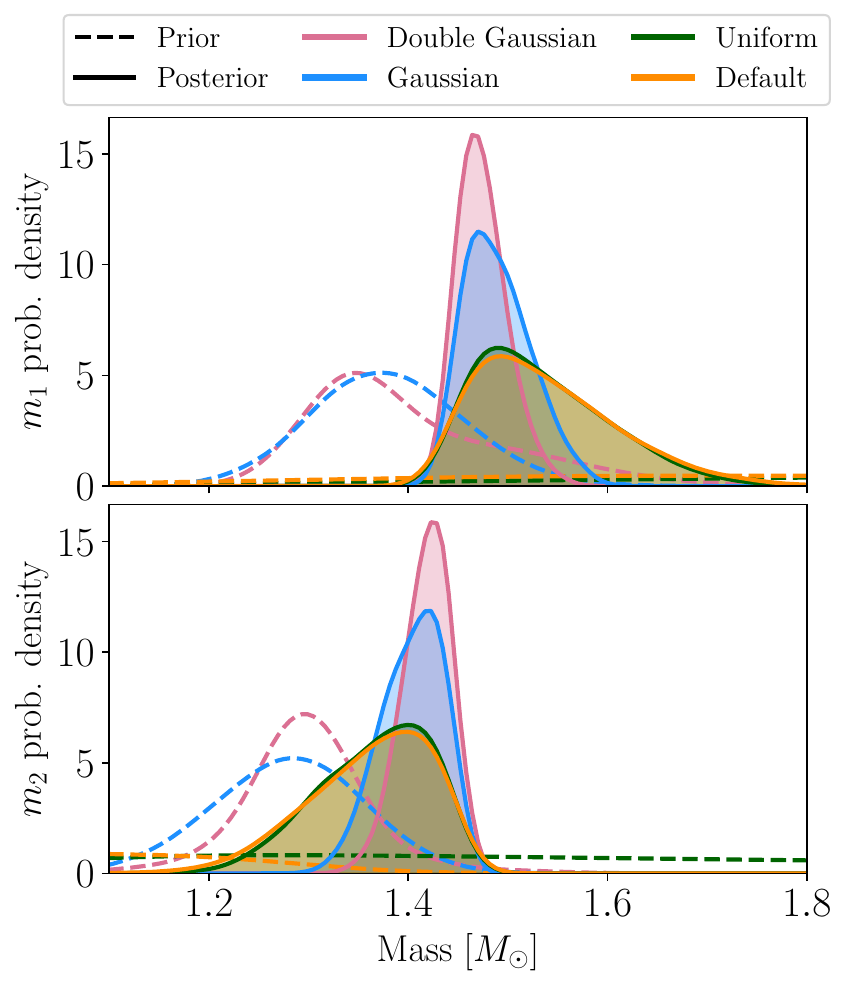}
    \caption{Comparison between priors (dashed lines) from populations and corresponding posteriors (solid lines) of the source-frame component masses, for the four mass priors considered.}
    \label{fig: populations and component masses KDEs}
\end{figure}

Comparing the posteriors obtained with different mass priors to the population models that determine those priors (cf.~Sec.~\ref{ssec: GW PE priors}) can hint at the underlying \ac{BNS} population.
Figure~\ref{fig: populations and component masses KDEs} shows this comparison visually with the prior and posterior distributions for the source-frame component masses in dashed and solid lines, respectively. The different colors refer to the different population models that determined the priors. As a caveat, we note that the search performed in Ref.~\cite{Niu:2025nha} assumed a double-Gaussian population model.

To quantify the degree of similarity between the various population-prior and posterior distributions, we use the \ac{JSD}~\cite{Lin:1991zzm}.
A \ac{JSD} value of $0$ bits indicates that the two distributions are the same, while a value of $1$ bit signifies they are completely different.
The \ac{JSD} values are given in Table~\ref{tab: JSD for masses and populations} in Appendix~\ref{app: JSD table}.

For both the primary and secondary masses, we observe that the distribution pair with the lowest \ac{JSD}, indicating the most consistent distributions, does not correspond to the scenario in which the population prior aligns with the prior used to obtain that posterior. This suggests that the candidate is informative and contributes measurable information beyond the prior.

For both masses, we find that all posteriors obtained with various priors are most consistent with the double Gaussian prior.
Indeed, the \ac{JSD} value between any posterior is the lowest when compared against the double Gaussian mass prior for both $m_1^{\rm{src}}$ and $m_2^{\rm{src}}$, indicating that this prior distribution is most similar to the posterior.
\footnote{We note that this conclusion remains also valid when using a double Gaussian with $\mathcal{N}(1.34\,\Msun, (0.08 \,\Msun)^2)$ and a relative weight of $0.65$, and $\mathcal{N}(1.80\,\Msun, (0.21\,\Msun)^2)$ with a relative weight of $0.35$ as in Ref.~\cite{Alsing:2017bbc}.}
However, it should be emphasized that the GW231109 candidate was identified in Ref.~\cite{Niu:2025nha} through a search that used a double Gaussian prior on the \ac{BNS} masses, which raised the significance of the candidate.
Therefore, the analysis presented in this section merely serves as a consistency check for the search performed in Ref.~\cite{Niu:2025nha} and should not be taken as an inference of the population of \acp{NS}, for which a more consistent treatment as well as other \ac{NS} observations should be taken into account (see, e.g., Refs.~\cite{Landry:2021hvl,Li:2021oyk,Biswas:2024hja,Golomb:2024lds,LIGOScientific:2025pvj}).

\subsection{Implications for the equation of state}\label{sec:eos}

Given the potential \ac{BNS} nature of the GW231109 candidate, we assess the possible implications it can have for the \ac{EOS}.
We use the same \ac{EOS} parametrization as Ref.~\cite{Koehn:2024set}, consisting of a fixed crust~\cite{Douchin:2001sv}, the metamodel parametrization~\cite{Margueron:2017eqc,Margueron:2017lup,Somasundaram:2020chb} for densities around the nuclear saturation density ($\nsat \equiv 0.16~\rm{fm}^{-3}$) up to a breakdown density~\cite{Tews:2018iwm}, above which the \ac{EOS} is extended with a piecewise speed-of-sound scheme~\cite{Tews:2018iwm,Greif:2018njt,Tews:2019cap,Somasundaram:2021clp}.
The \ac{EOS} parametrization is sampled with \textsc{Jester}~\cite{Wouters:2025zju}, which accelerates the inference through GPU acceleration~\cite{frostig2018compiling,kidger2022neuraldifferentialequations}.
Contrary to Ref.~\cite{Wouters:2025zju}, we use the sequential Monte Carlo sampler~\cite{del2006sequential,chopin2020introduction} implemented in \textsc{blackjax}~\cite{cabezas2024blackjax} as our stochastic sampler, evolving $5000$ particles using an adaptive temperature ladder with a target effective sample size of $90\%$ and $10$ MCMC steps per temperature step.

During sampling, we incorporate a likelihood term that disfavors \acp{EOS} predicting a \ac{TOV} mass below the mass of the heaviest observed \acp{PSR}, namely, PSR J1614-2230~\cite{Demorest:2010bx,Shamohammadi:2022ttx} and \ac{PSR} J0740+6620~\cite{Fonseca:2021wxt} (see also Refs.~\cite{NANOGrav:2019jur,Riley:2021pdl,Salmi:2022cgy,Miller:2021qha,NANOGrav:2023hde,Dittmann:2024mbo,Salmi:2024aum}).
Throughout this work, we refer to this constraint as `heavy \acp{PSR}'.

Additionally, for the \ac{BNS} events and the GW231109 trigger considered in this work, we use the marginalized 4-dimensional posteriors on the source-frame component masses and tidal deformabilities to approximate the pseudo-likelihood function for the \ac{EOS} parameters.
The marginal posteriors are approximated by a normalizing flow~\cite{decao2019blockneuralautoregressiveflow,ward2023flowjax}: see Ref.~\cite{Wouters:2025zju} for details.

Table~\ref{tab: R14 main results} shows the posterior median value and $90\%$ credible intervals of the radius of a $1.4\,\Msun$ \ac{NS}.
The constraints on the \ac{TOV} mass and the pressure at $3\nsat$, as well as the results obtained from other prior choices for GW231109, are given by Tab.~\ref{tab: eos_parameters} in Appendix~\ref{app: more EOS results}.
As expected, low-\ac{SNR} events such as GW190425 and the candidate trigger GW231109 do not impose any noticeable constraints on the \ac{EOS}, and the constraints are driven by GW170817 due to its high \ac{SNR} (as already noted by Refs.~\cite{DelPozzo:2013ala,Lackey:2014fwa,HernandezVivanco:2019vvk}).
Compared to the joint inference of GW170817 and GW190425, the inclusion of the GW231109 candidate implies a slightly lower median value for $R_{1.4}$, although it has a negligible impact on the width of the $90\%$ credible interval, again due to the lower \ac{SNR}.

It should be emphasized that the \ac{EOS} inference treats all three \ac{BNS} candidates on an equal footing, and therefore neglects the uncertainty in the nature of the GW231109 signal.
Together with the weak constraints obtained from this low-\ac{SNR} candidate, we conclude that no meaningful conclusions can be drawn regarding the \ac{EOS} by incorporating this candidate.

\begin{table}[t]
    \centering
    \caption{Posterior on $R_{1.4}$ obtained from the constraints discussed in Sec.~\ref{sec:eos}, denoted by the median with $90\%$ credible intervals as uncertainty.
    All inferences with \ac{GW} data also include the constraints from heavy \acp{PSR}.
    }
    \input{eos_r14_table}
    \label{tab: R14 main results}
\end{table}

\subsection{Estimating the remnant fate}\label{sec:remn}

Depending on the properties of the system, the remnant formed during the merger process can have different lifetimes before collapsing into a \ac{BH} or even remain stable~\cite{Hotokezaka:2011dh,Hotokezaka:2013iia,Sarin:2020gxb,Dietrich:2020eud,Bernuzzi:2020tgt}. Typically, for heavier masses, the system promptly collapses to a \ac{BH}. Alternatively, a differentially-rotating hyper-massive neutron star (HMNS) can be formed, with a possibly higher mass than the maximum mass sustained by a uniformly rotating \ac{NS}. This HMNS survives for a few milliseconds before different mechanisms dissipate the differential rotation, either causing it to collapse to a \ac{BH} or, in the case of slightly lower masses, to form a uniformly rotating supra-massive \ac{NS} (SMNS) that survives up to $\mathcal{O}(1 \, {\rm s})$ before collapsing. For very low-mass systems, the mass of the remnant produced by the merger can be lower than the \ac{TOV} mass, and therefore a stable \ac{NS} is formed.

To predict the remnant expected for this trigger, we use the classifiers developed in Ref.~\cite{Puecher:2024dhl}, based on gradient boost decision trees~\cite{friedman2001greedy,FRIEDMAN2002367}.
The classifiers are trained on a set of $\sim 400$ publicly available numerical relativity simulations (with $10\%$ used as validation), whose outcome is classified based on the time between merger and collapse to a black hole.
The input features are quantities that can be measured from the inspiral GW signal, without the need for any postmerger observation.
The models' hyperparameters are optimized through \texttt{GridSearchCV}~\cite{grid_search} and early stopping.
More details about the classifiers, the training procedure, and their accuracy can be found in Ref.~\cite{Puecher:2024dhl}.
In particular, three different classifiers are available: \emph{Classifier A} distinguishes between prompt collapse to \ac{BH} ($p_{\rm \textsc{pcbh}}$) or formation of a \ac{NS} remnant ($p_{\rm \textsc{rns}}$); \emph{Classifier B} further distinguishes the \ac{NS} remnant scenario in formation of a HMNS ($p_{\rm \textsc{hmns}}$) or a remnant (which could be a HMNS, SMNS, or stable NS) that survives more than $25 \, \rm{ms}$ ($p_{\rm \textsc{nc}}$); \emph{Classifier C}, in which the HMNS class is further divided into short-lived ($p_{\rm \textsc{short}}$), i.e., for which the HMNS collapses to a BH in a time $2 \, {\rm ms} < \tau_{\rm BH} < 5 \, {\rm ms}$, and long-lived ($p_{\rm \textsc{long}}$), i.e., with $\tau_{\rm BH} > 5 \, {\rm ms}$.
All three classifiers predict the remnant based on the values of source parameters that can be inferred from the inspiral signal: the total mass, the mass-weighted tidal deformability $\tilde{\Lambda}$, the mass ratio $q$, and the effective spin $\chi_{\rm eff}$.

Table~\ref{tab:remn} shows the probabilities for the different kinds of remnants based on the posteriors both from our fiducial run with EOS sampling and low-spin prior ($a_i < 0.05$), and the EOS-sampling analysis with larger spin prior ($a_i < 0.4$).
We do not apply the classifier to posteriors obtained with uniform priors on the tidal deformabilities or to those assuming the \ac{QUR}, since the candidate’s low \ac{SNR} leads to a poorly constrained $\tilde{\Lambda}$, producing samples with nonphysical combinations of masses and tidal deformabilities.
In both cases and for all three classifiers, the preferred scenario is a prompt collapse to a \ac{BH}, although there is a non-negligible probability for the formation of an HMNS.
The prompt-collapse probability is higher for the high-spin analysis, consistent with the overall higher mass inferred (see Figure~\ref{fig: GW PE cornerplots}).
In case an HMNS was formed, it was most likely short-lived, i.e., it collapsed within 5~ms from merger.

\begingroup
\renewcommand*{\arraystretch}{2}
\setlength{\tabcolsep}{8pt}
\begin{table}[t]
\caption{Probabilities of the different merger outcomes using the three classifiers developed in Ref.~\cite{Puecher:2024dhl} applied to the posteriors obtained while sampling the \ac{EOS}, for both low-spin and high-spin priors. The numbers between parenthesis show the confidence of the prediction, computed as the mean of the samples prediction confidence (see Ref.~\cite{Puecher:2024dhl} for details).}
\begin{flushleft}
\begin{tabular}{l|cccc}
\hline\hline
\textbf{Classifier A} & $p_{\textsc{pcbh}}$ & $p_{\textsc{rns}}$ & & \\
\hline
Low spin ($91.7\%$) & $57.4\%$ & $42.6\%$ & & \\
High spin ($94.8\%$)& $79.1\%$ & $20.9\%$ & & \\
\hline
\textbf{Classifier B} & $p_{\textsc{pcbh}}$ & $p_{\textsc{HMNS}}$ & $p_{\textsc{NC}}$ & \\
\hline
Low spin  ($58.2\%$)  & $57.9\%$ & $42.0\%$ & $0.1\%$ & \\
High spin ($78.1\%$) & $79.1\%$ & $20.8\%$ & $0.1\%$ &\\
\hline
\textbf{Classifier C} & $p_{\textsc{pcbh}}$ & $p_{\textsc{short}}$ & $p_{\textsc{long}}$ & $p_{\textsc{nc}}$ \\
\hline
Low spin  ($58.3\%$) & $62.8\%$ & $35.0\%$ & $2.1\%$ & $0.0\%$ \\
High spin ($79.5\%$)& $83.1\%$ & $14.3\%$ & $1.8\%$ & $0.7\%$ \\
\hline\hline
\end{tabular}
\label{tab:remn}
\end{flushleft}
\end{table}
\endgroup

\subsection{Estimating possible kilonova lightcurves}\label{ssec: KN LC}

Given the non-negligible probability of an HMNS formation, we predict the corresponding kilonova light curves to assess what \ac{EM} counterpart this candidate would have produced.
For this purpose, we use the nuclear-physics and multi-messenger astrophysics framework
\textsc{nmma}~\cite{Pang:2022rzc}, and employ the \texttt{Bu2019lm} model~\cite{Dietrich:2020efo}, together with the posteriors from our fiducial run with \ac{EOS} sampling.
The \texttt{Bu2019lm} model is built using \texttt{POSSIS}~\cite{Bulla:2019muo,Bulla:2022mwo}, a radiative transfer code simulating photon packets diffusing out of the ejected material.
Both the dynamic ejecta mass \(M^{\rm ej}_{\rm dyn}\) and the disk mass $M_{\rm disk}$ are estimated using the relations presented in Ref.~\cite{Pang:2022rzc}.
The wind ejecta mass $M^{\rm ej}_{\rm wind}$ is assumed to be 30\% of the disk mass~\cite{Fujibayashi:2017puw,Lund:2024fjk}.
The opening angle $\Phi$ between the lanthanide-rich and lanthanide-poor dynamical ejecta components is varied within the range of $[15^{\circ}, 75^{\circ}]$.
The estimated ejecta masses are $\log_{10}M^{\rm ej}_{\rm dyn} / \Msun = -2.19^{+0.34}_{-0.13}$ and $\log_{10}M^{\rm ej}_{\rm wind} / \Msun = -1.27^{+0.21}_{-0.02}$.
The quoted values represent the median along with the $90\%$ credible interval as uncertainty.

\begin{figure}
\centering
\includegraphics[width=\linewidth]{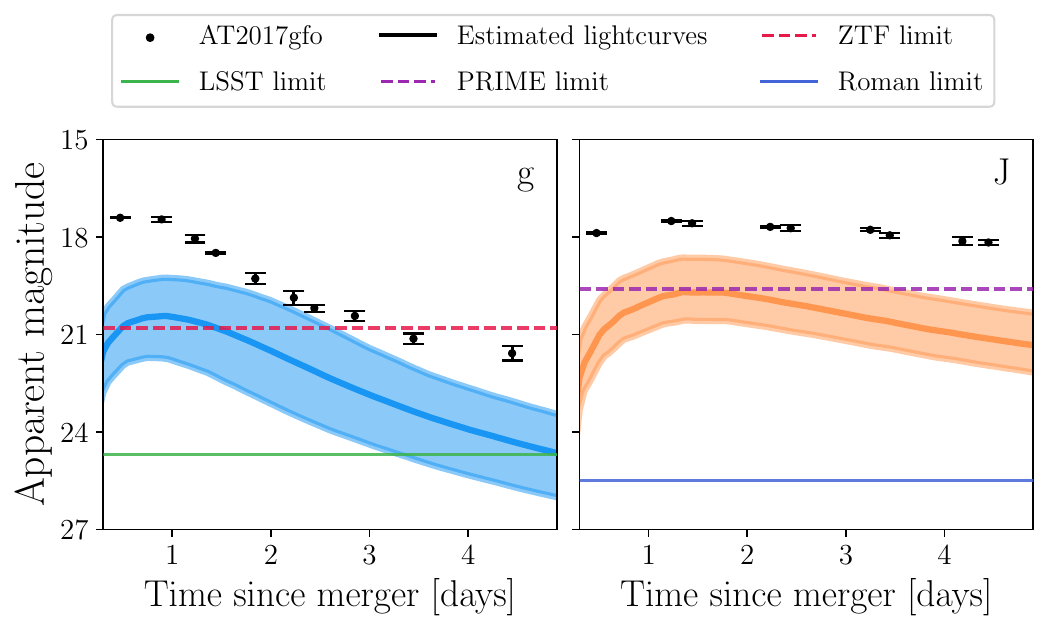}
\caption{Lightcurves showing the median values along with the 90\% credible interval, measured in AB magnitudes across various photometric bandpasses, compared with data from AT2017gfo~\cite{LIGOScientific:2017pwl,Andreoni:2017ppd,Coulter:2017wya,Cowperthwaite:2017dyu,Drout:2017ijr,Lipunov:2017dwd,Shappee:2017zly,Tanvir:2017pws,Troja:2017nqp,J-GEM:2017tyx,Villar:2017wcc}.~The horizontal lines show limiting magnitudes of various instruments, taken from Ref.~\cite{Chase:2021ood}.}
\label{fig: lc}
\end{figure}

The estimated lightcurves, along with the observations from AT2017gfo~\cite{LIGOScientific:2017pwl,Andreoni:2017ppd,Coulter:2017wya,Cowperthwaite:2017dyu,Drout:2017ijr,Lipunov:2017dwd,Shappee:2017zly,Tanvir:2017pws,Troja:2017nqp,J-GEM:2017tyx,Villar:2017wcc}, are presented in Figure~\ref{fig: lc}.
The lightcurves shown represent the median values, along with the $90\%$ credible interval, measured in AB magnitudes across various photometric bandpasses.
It is evident that for this candidate, at a distance of around $165$~Mpc, the kilonova lightcurves are significantly dimmer than for AT2017gfo.
Based on the limiting magnitudes reported in Ref.~\cite{Chase:2021ood} and on predictions from our lightcurve model, we find that ZTF~\cite{ZTF} and PRIME~\cite{Sumi_2025} might have been able to detect the peak of the lightcurve, although this is dependent on the uncertainties in the trigger's source parameters.
The LSST observatory~\cite{LSST:2008ijt} and the Roman Space Telescope~\cite{Spergel:2015sza}, on the other hand, certainly have the sensitivity to detect the transient and track the lightcurve evolution over several days.

Reference~\cite{Li:2025rmj} performed a search for an \ac{EM} counterpart of GW231109 with archival data and reported one candidate, AT2023xqy, to be in good agreement with the trigger time and distance reported by the \ac{GW} pipelines.
However, given that AT2023xqy peaks roughly $14$ days after the \ac{GW} trigger, our results make it challenging to interpret this transient as a plausible kilonova counterpart of GW231109.

\section{Injections with third-generation detectors}
\label{sec:inj}

To evaluate the potential implications on the \ac{EOS} from a source like GW231109, we simulate a similar source detected by \ac{3G} detectors.
These detectors are anticipated to observe up to $\mathcal{O}(10^4)$ \ac{BNS} events annually, with around $\mathcal{O}(10^3)$ of those having an \ac{SNR} greater than $100$~\cite{ET:2019dnz,Abac:2025saz}.
Two detector configurations are considered.
First, we assume \ac{ET} in its triangular xylophone configuration (ET-$\Delta$), with $10$~km arms and located in the EMR region~\cite{et_psd}.
Second, we consider a $2$L design (ET-2L)~\cite{Branchesi:2023mws}, with one L placed in the EMR region and one in Sardinia, and both having $15$~km arms.
Additionally, for both designs, we consider \ac{ET} in a network with \ac{CE}, where \ac{CE} is assumed to be an L-shaped detector with 40~km arm-length located at the current LIGO-Hanford site~\cite{CE-T2000017-v8}.

The injected parameters are the median values of the posteriors obtained in our fiducial analysis with \ac{EOS} sampling (i.e., the low-spin prior run in the right panel of Figure~\ref{fig: GW PE cornerplots}).
However, given the poor constraints on the spin angle parameters, the median values would correspond to highly precessing systems, which are not expected in the known \ac{BNS} population and could induce biases in the analysis due to the models' limited calibration accuracy in this parameter region.
Therefore, we simulate a system with aligned spins.
The injected \ac{EOS}, from which the tidal deformabilities are computed, is the maximum likelihood \ac{EOS} inferred using only heavy \acp{PSR}, as described in Sec.~\ref{sec:eos}.
The simulated signals yield \ac{SNR} values of $132$ and $367$ for the ET-$\Delta$ and ET-$\Delta$+CE networks, respectively, and $167$ for ET-$2$L and $381$ for ET-$2$L+CE.

The recovery uses the \texttt{IMRPhenomXP\_\allowbreak NRTidalv3} waveform, without assuming aligned spins, to be consistent with the previous analyses, and with the default mass, low-spin, and uniform tidal deformability priors.
The starting frequencies for the analysis are set to $5$ Hz for \ac{ET} and $10$ Hz for \ac{CE}.\footnote{Being planned to be built underground, ET is expected to gain sensitivity also below 10~Hz.}
In this work, we do not consider effects from the rotation of the Earth.

\begin{figure}[t]
     \centering
     \includegraphics[width=0.99\columnwidth]{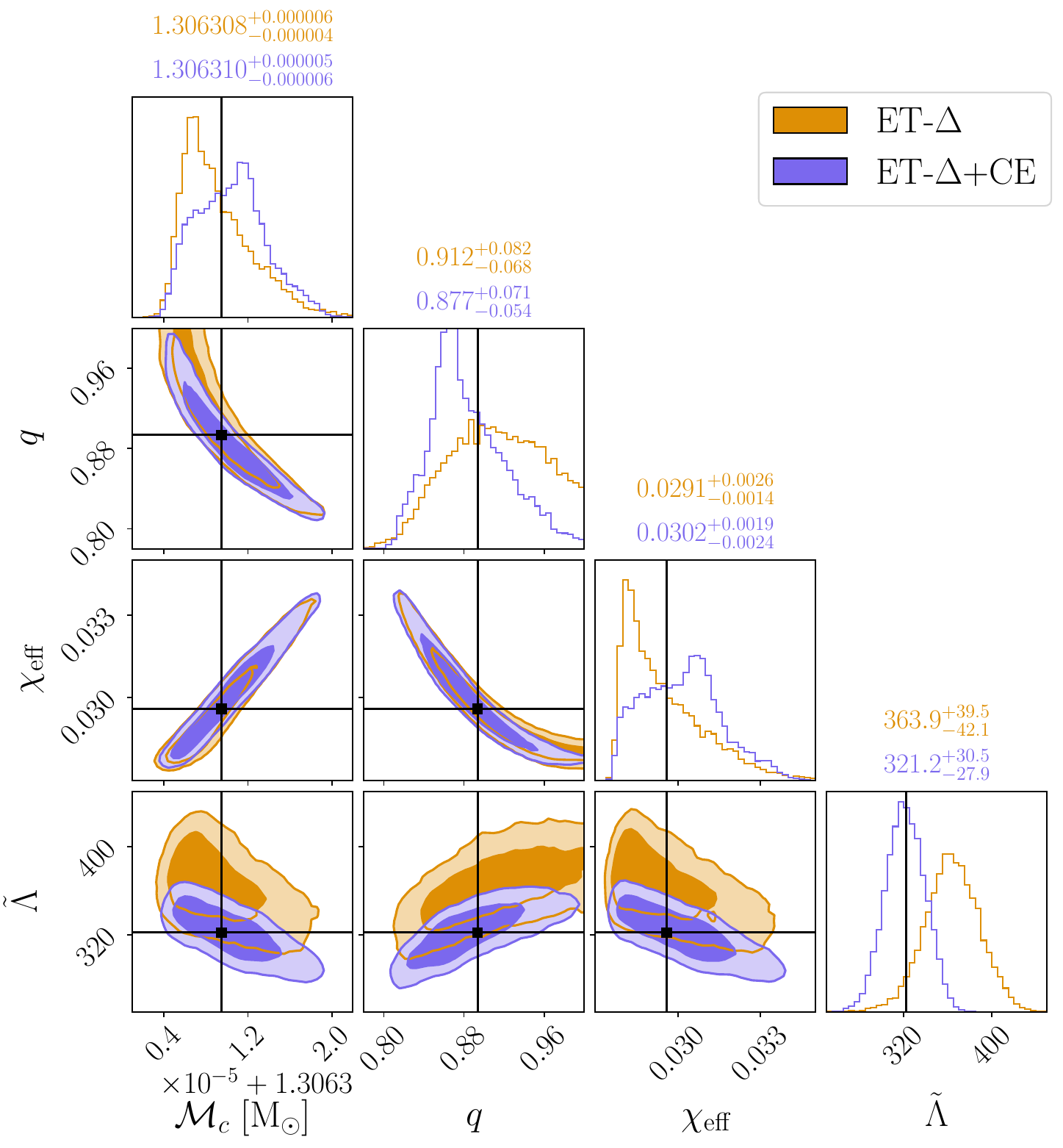}
     \caption{Posterior on parameters from a simulated source similar to GW231109 observed with ET-$\Delta$ and a network of ET-$\Delta$ with CE. The black lines indicate the injected values. The median values of the recovered parameters, together with the $90\%$ credible intervals, are reported above the marginalized posterior distributions.}
     \label{fig: ET and ET+CE GW PE results}
 \end{figure}

Figure~\ref{fig: ET and ET+CE GW PE results} shows the resulting posteriors for chirp mass $\mathcal{M}_c$, mass ratio $q$, effective spin $\chi_{\rm eff}$, and mass-weighted tidal deformability $\tilde{\Lambda}$, for the ET-$\Delta$ design.
All parameters are well recovered with an improved accuracy compared to the posteriors from Figure~\ref{fig: GW PE cornerplots}, due to the higher \ac{SNR}.
As expected, the uncertainties are further reduced when considering \ac{ET} in a network with \ac{CE}.
Similar observations hold for ET-$2$L and are shown in Fig.~\ref{fig: ET2L and ET2L+CE GW PE results} in Appendix~\ref{sec: appendix ET2L}.

\begin{figure}[t]
    \centering
    \includegraphics[width=0.99\columnwidth]{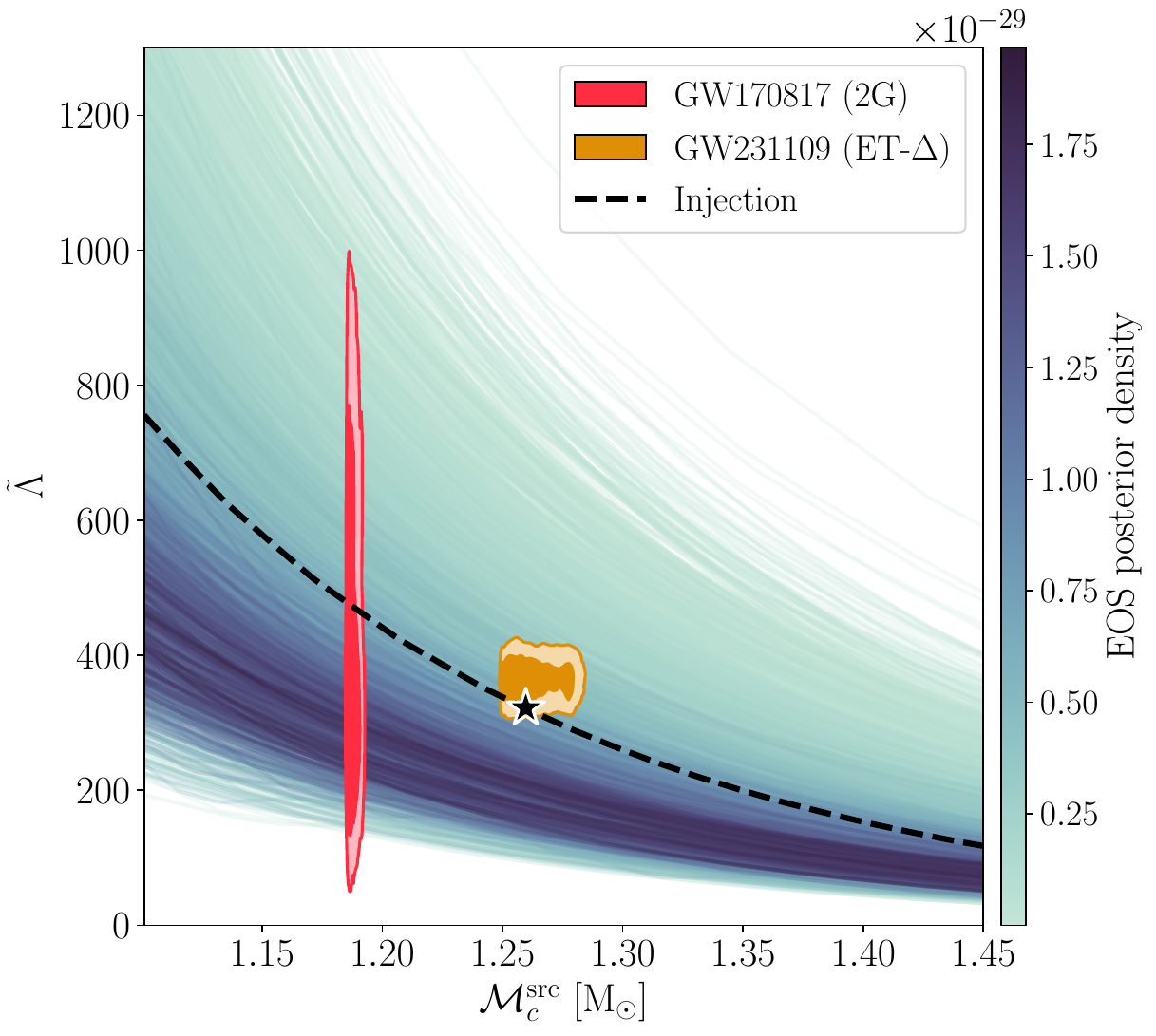}
    \caption{Posterior samples on source-frame chirp mass and $\tilde{\Lambda}$ for GW170817 (pink) and the simulated signals (orange shades), together with posterior \ac{EOS} samples constrained by GW170817, color-coded by the posterior probability.
    The \ac{EOS} used for the injections, as well as the simulated signal, are shown in black.
    The dark (light) shades represent $68\%$ ($95\%$) credible areas.
    }
    \label{fig: EOS curves and Mc source and lambda tilde}
\end{figure}

Figure~\ref{fig: EOS curves and Mc source and lambda tilde} shows the posteriors on the source-frame chirp mass $\mathcal{M}^{\rm{src}}$ and mass-weighted tidal deformabilities for the simulated source as detected by 3G detectors, compared to GW170817.
The uncertainty in the source-frame chirp mass is larger for the simulated signals compared to GW170817, despite the detector-frame chirp mass being well-measured thanks to the high \ac{SNR}.
This is because the redshift values in the posteriors of the simulated signals have a larger spread at the larger distance of $168$ Mpc.
We also show posterior samples from the \ac{EOS} inference outlined in Sec.~\ref{sec:eos} using GW170817 as a constraint, also shown in Table~\ref{tab: R14 main results}, as well as the \ac{EOS} used to compute the tidal deformabilities for the injection.
The curves are plotted assuming an equal-mass system for visualization.
Due to the tidal deformabilities being well measured at this high \ac{SNR}, the simulated events are more informative for the \ac{EOS} compared to GW170817.

\begin{table}[t]
    \centering
    \caption{Posterior on $\Lambda_{1.4}$ for the 3G injections with different detector networks, denoted by the median with $90\%$ credible intervals as uncertainty.
    The injected value is $\Lambda_{1.4}^{\rm{inj}} = 399$.}
    \input{3G_lambdas}
    \label{tab: R14 main results 3G}
\end{table}

The obtained \ac{GW} source parameters are then used to constrain the \ac{EOS}, as described in Sec.~\ref{sec:eos}.
The resulting constraints on the tidal deformability of a $1.4 \ \Msun$ \ac{NS} ($\Lambda_{1.4}$) are shown in Tab.~\ref{tab: R14 main results 3G}, with the injected value at $\Lambda_{1.4}^{\rm{inj}} = 399$.
The median values of the \ac{EOS} posteriors are consistent with the $\tilde{\Lambda}$ measurements from the \ac{GW} posterior: ET-$\Delta$ slightly overestimates $\tilde{\Lambda}$ and therefore also $\Lambda_{1.4}$, while the networks including CE have their $\Lambda_{1.4}$ posteriors peak close to the injected value.
The uncertainties are, as expected, purely driven by the \ac{SNR} of the detection, going from $\sim12\%$ relative uncertainty (for ET-$\Delta$) to $\sim7.5\%$ (for ET-2L+CE).
While ET-2L generally yields narrower \ac{EOS} posteriors than ET-$\Delta$ for this single source, as a result of the increased \ac{SNR}, Ref.~\cite{Iacovelli:2023nbv} noted that the \ac{EOS} extraction of both designs are equally performant when considering a population of a few hundred detections.

We note that our projection is optimistic, since, apart from precession effects, the injection and recovery use the same waveform, such that effects from waveform systematics can be ignored.
This will become important for signals with an \ac{SNR} above $\gtrsim 80$~\cite{Gamba:2020wgg}, when mismodelling effects in the \ac{BNS} inspiral become non-negligible~\cite{Pratten:2021pro}, or when combining information from a few tens of \ac{BNS} mergers~\cite{Kunert:2021hgm}.

\section{Conclusions}
\label{sec:conclusions}

In this work, we investigated GW231109\_235456 (abbreviated GW231109), a sub-threshold \ac{BNS} candidate identified by Ref.~\cite{Niu:2025nha} in LIGO-Virgo-KAGRA's O4a observing run.
Although the astrophysical nature of the signal is still uncertain, we provide in-depth follow-up investigations of this candidate.
By using state-of-the-art \ac{BNS} waveform models, various prior choices, and extending the frequency range of the analysis to $2048$ Hz, we aim to provide a comprehensive discussion of the \ac{GW} trigger and its potential implications.
Due to the low \ac{SNR} of the signal, prior assumptions noticeably influence the final posterior distributions.

By using different priors on the source-frame component masses, we find that the source is most consistent with the double Gaussian population model.

We additionally constrain the \ac{EOS} from GW231109, finding that the \ac{EOS} is poorly constrained from GW231109 alone, due to the low \ac{SNR} of the candidate.
Similarly, when combined with GW170817 and GW190425, the GW231109 candidate has a negligible impact on the \ac{EOS}.

Furthermore, we predict the fate of the remnant formed after the merger using machine-learning classifiers trained on numerical-relativity simulations of \ac{BNS} mergers from Ref.~\cite{Puecher:2024dhl}.
We find that the merger most likely led to a prompt collapse to a \ac{BH} and, if an HMNS was formed first, it was likely short-lived, i.e., collapsed to a \ac{BH} around $2$--$5$~ms after merger.
We also predict the lightcurves from a potential kilonova counterpart that would originate from the merger.
While the lightcurves are dimmer than AT2017gfo due to the larger distance, instruments such as LSST or the Roman Space Telescope could have been able to detect an electromagnetic counterpart of this candidate.

Finally, we use the inferred source properties of GW231109 to simulate a signal with similar properties as observed by \ac{ET}, and a network of \ac{ET} and \ac{CE}.
Due to the increased \ac{SNR}, the source properties are better constrained, which yields relative uncertainties on $\Lambda_{1.4}$ of about $10\%$, depending on the specific designs for ET and network of detectors considered.

\section*{Data availability}

Code and data to reproduce the results in this manuscript is available at Ref.~\cite{paper_repo}.

\begin{acknowledgement}
We thank Lami Suleiman, the LIGO-Virgo-KAGRA extreme matter community, Mick Wright, Michael Williams, and Luca Negri for fruitful discussions and feedback that led to the improvement of this work.
T.W. is supported by the research program of the Netherlands Organization for Scientific Research (NWO) under grant number OCENW.XL21.XL21.038.
A.P., T.D. acknowledge funding from the EU Horizon under ERC Starting Grant, no.\ SMArt-101076369.
P.T.H.P. is supported by the research program of the Netherlands Organization for Scientific Research (NWO) under grant number VI.Veni.232.021.
We thank SURF (www.surf.nl) for the support in using the National Supercomputer Snellius under project number EINF-14622.
The computations were performed on the DFG-funded research cluster Jarvis at the University of Potsdam (INST 336/173-1; project number: 502227537).
Views and opinions expressed are those of the authors only and do not necessarily reflect those of the European Union or the European Research Council. Neither the European Union nor the granting authority can be held responsible for them. This research has made use of data or software obtained from the Gravitational Wave Open Science Center (gwosc.org), a service of the LIGO Scientific Collaboration, the Virgo Collaboration, and KAGRA. This material is based upon work supported by NSF's LIGO Laboratory which is a major facility fully funded by the National Science Foundation, as well as the Science and Technology Facilities Council (STFC) of the United Kingdom, the Max-Planck-Society (MPS), and the State of Niedersachsen/Germany for support of the construction of Advanced LIGO and construction and operation of the GEO600 detector. Additional support for Advanced LIGO was provided by the Australian Research Council. Virgo is funded, through the European Gravitational Observatory (EGO), by the French Centre National de Recherche Scientifique (CNRS), the Italian Istituto Nazionale di Fisica Nucleare (INFN) and the Dutch Nikhef, with contributions by institutions from Belgium, Germany, Greece, Hungary, Ireland, Japan, Monaco, Poland, Portugal, Spain. KAGRA is supported by Ministry of Education, Culture, Sports, Science and Technology (MEXT), Japan Society for the Promotion of Science (JSPS) in Japan; National Research Foundation (NRF) and Ministry of Science and ICT (MSIT) in Korea; Academia Sinica (AS) and National Science and Technology Council (NSTC) in Taiwan.
T.W. acknowledges the use of generative AI (Claude Sonnet 4.5) to proofread the manuscript.
All AI-generated outputs were carefully reviewed and edited by T.W. to ensure accuracy.
\end{acknowledgement}

\appendix

\section{Jensen-Shannon divergences}\label{app: JSD table}

Table~\ref{tab: JSD for masses and populations} shows the \ac{JSD} values between the prior and posterior distributions on which the discussion presented in Sec.~\ref{ssec: populations} is based.

\begingroup
\renewcommand*{\arraystretch}{2}
\begin{table*}[t]
\centering
\resizebox{\textwidth}{!}{\input{JSD_tabular}}
\caption{JSD values (in bits) between the prior and posterior distributions of the source-frame component masses, for the various mass priors described in Sec.~\ref{ssec: GW PE priors} and using the default low-spin, uniform in $\Lambda_i$ priors. Bold values denote the lowest \ac{JSD} within each row.}
\label{tab: JSD for masses and populations}
\end{table*}
\endgroup

\section{Constraints on \ac{EOS} quantities}\label{app: more EOS results}

In Table~\ref{tab: eos_parameters} we show the constraints on the \ac{EOS} by summarizing a few quantities of interest.
In particular, we show the \ac{TOV} mass, $\MTOV$, the radius of a $1.4\,\Msun$ \ac{NS}, $R_{1.4}$, and the pressure at $3\nsat$, $p(3\nsat)$, for the various constraints discussed in Sec.~\ref{sec:eos}.
The inference results using information from a \ac{BNS} candidate use the same default prior as defined in Sec.~\ref{ssec: GW PE priors}, i.e., using the default mass prior, low-spin prior, and sampling $\Lambda_i$ uniformly in the range $[0, 5000]$ for all \ac{BNS}.
When considering GW231109 alone, we infer the \ac{EOS} by changing one aspect of these prior choices, as denoted by the brackets and explained in detail in Sec.~\ref{ssec: GW PE priors}.
In particular, the label \texttt{XAS} denotes inferring the \ac{EOS} from the posterior obtained when using the aligned-spin waveform approximant \texttt{IMRPhenomXAS\_\allowbreak NRTidalv3} in the recovery.

\begin{table*}[t]
\centering
\caption{Constraints on $\MTOV$, $R_{1.4}$ and $p(3\nsat)$. For GW231109, we show the constraints for various prior choices as described in Sec.~\ref{ssec:pe_res}. All constraints using \ac{BNS} mergers additionally enforce the heavy pulsar constraint.
All values denote 90\% credible intervals.
}
\label{tab: eos_parameters}
\resizebox{\textwidth}{!}{\input{eos_parameters_table}}
\end{table*}

\section{GW posteriors for ET-2L}\label{sec: appendix ET2L}

\begin{figure}[t]
     \centering
     \includegraphics[width=0.99\columnwidth]{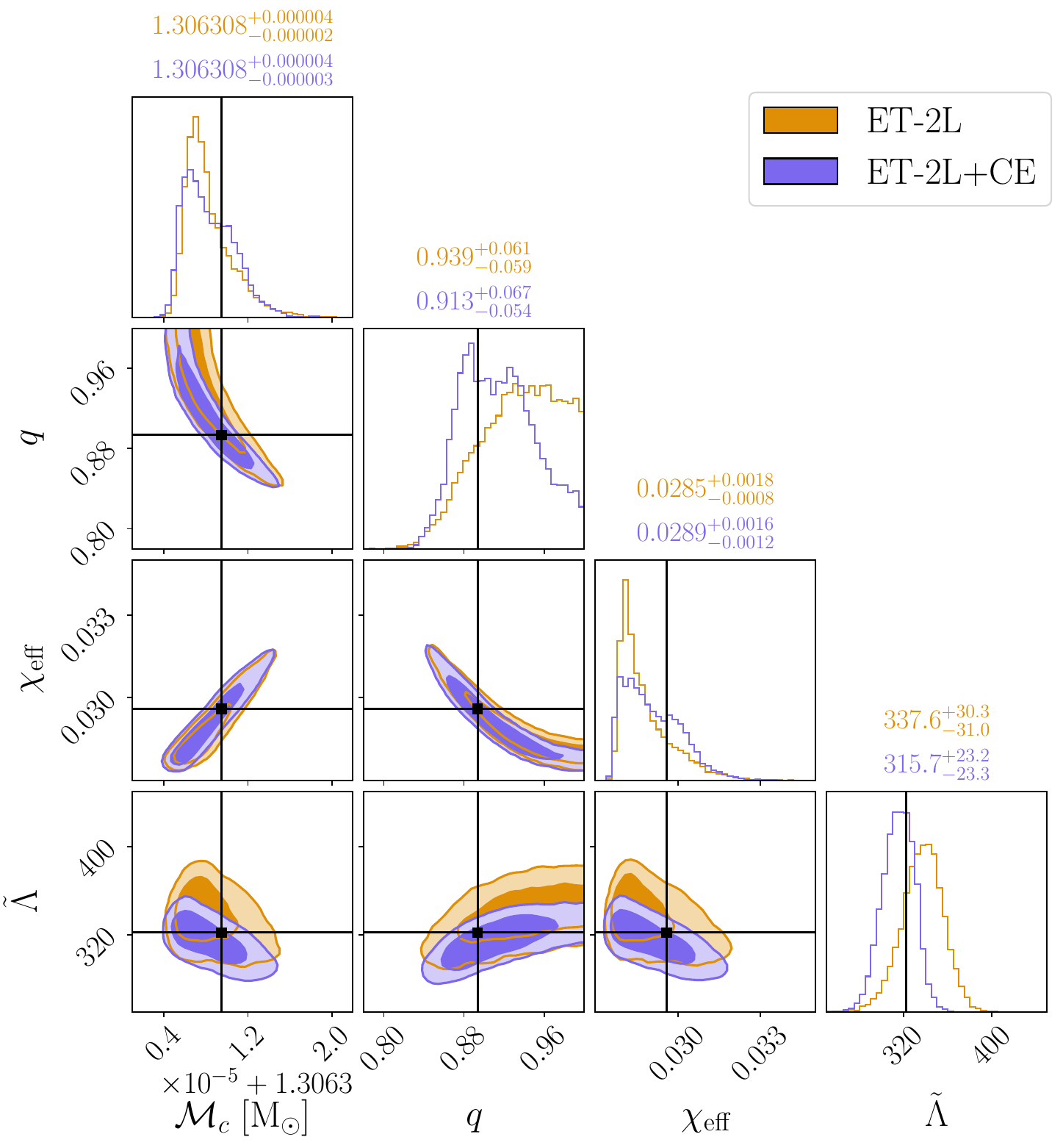}
     \caption{Posterior on parameters from a simulated source similar to GW231109 observed with ET-2L and a network of ET-2L with CE. The black lines indicate the injected values. The median values of the recovered parameters, together with the $90\%$ credible intervals, are reported above the marginalized posterior distributions.}
     \label{fig: ET2L and ET2L+CE GW PE results}
 \end{figure}

In Fig.~\ref{fig: ET2L and ET2L+CE GW PE results}, we show similar posterior distributions from the \ac{GW} inference as shown in Fig.~\ref{fig: ET and ET+CE GW PE results}, but instead showing the results for the $2$L design of \ac{ET}.

\begin{acronym}
    \acro{AD}[AD]{automatic differentiation}
    \acro{JIT}[JIT]{just-in-time}
    \acro{PE}[PE]{parameter estimation}
    \acro{ROQ}[ROQ]{reduced order quadrature}
    \acro{MCMC}[MCMC]{Markov chain Monte Carlo}
    \acro{LVK}[LVK]{LIGO-Virgo-KAGRA}
    \acro{3G}[3G]{third-generation}
    \acro{PSR}[PSR]{pulsar}
    \acro{GW}[GW]{gravitational wave}
    \acrodefplural{GWs}{gravitational waves}
    \acro{QUR}[QUR]{quasi-universal relations}
    \acro{ET}[ET]{Einstein Telescope}
    \acro{CE}[CE]{Cosmic Explorer}
    \acro{EM}[EM]{electromagnetic}
    \acro{CBC}[CBC]{compact binary coalescences}
    \acro{NS}[NS]{neutron star}
    \acrodefplural{NSs}{neutron stars}
    \acro{KDE}[KDE]{kernel density estimate}
    \acro{NF}[NF]{normalizing flow}
    \acro{BBH}[BBH]{binary black hole}
    \acro{BNS}[BNS]{binary neutron star}
    \acro{BH}[BH]{black hole}
    \acro{NSBH}[NSBH]{neutron star-black hole}
    \acro{EOS}[EOS]{equation of state}
    \acro{EFT}[EFT]{effective field theory}
    \acro{chiEFT}[$\chi$EFT]{chiral effective field theory}
    \acro{NEP}[NEP]{nuclear empirical parameter}
    \acro{HIC}[HIC]{heavy-ion collision}
    \acrodefplural{NEPs}{nuclear empirical parameters}
    \acro{MM}[MM]{metamodel}
    \acro{CSE}[CSE]{speed-of-sound extension scheme}
    \acro{TOV}[TOV]{Tolman-Oppenheimer-Volkoff}
    \acro{JS}[JS]{Jensen-Shannon}
    \acro{JSD}[JSD]{Jensen-Shannon divergence}
    \acro{CPU}[CPU]{central processing unit}
    \acro{GPU}[GPU]{graphical processing unit}
    \acro{TPU}[TPU]{tensor processing unit}
    \acro{ML}[ML]{machine learning}
    \acro{SNR}[SNR]{signal-to-noise ratio}
    \acro{FAR}[FAR]{false alarm rate}
    \acro{PSD}[PSD]{power spectral density}
    \acro{NICER}[NICER]{Neutron star Interior Composition ExploreR}
\end{acronym}

\bibliography{references}{}
\bibliographystyle{unsrt}

\end{document}

%% file: eos_r14_table.tex
\begin{tabular}{l@{\hspace{1.5cm}}c}
\toprule\toprule
Dataset & $R_{1.4}$ [km] \\
\midrule
Prior & $13.41^{+1.99}_{-2.57}$ \\
\addlinespace
Heavy PSRs & $13.74^{+1.47}_{-1.48}$ \\
\addlinespace
GW170817 & $12.39^{+1.13}_{-1.34}$ \\
\addlinespace
GW190425 & $13.42^{+1.46}_{-1.40}$ \\
\addlinespace
GW231109 (default) & $13.35^{+1.70}_{-1.41}$ \\
\addlinespace
GW170817$+$GW190425 & $12.38^{+1.11}_{-1.27}$ \\
\addlinespace
GW170817$+$GW231109 & $12.36^{+1.21}_{-1.30}$ \\
\addlinespace
GW190425$+$GW231109 & $13.15^{+1.38}_{-1.42}$ \\
\addlinespace
GW170817$+$GW190425$+$GW231109 & $12.24^{+1.21}_{-1.20}$ \\
\addlinespace
\bottomrule\bottomrule
\end{tabular}

%% file: 3G_lambdas.tex
\begin{tabular*}{0.75\columnwidth}{@{\extracolsep{\fill}}lc}
\toprule\toprule
Network & $\Lambda_{1.4}$ \\
\midrule
ET-$\Delta$ & $447^{+54}_{-51}$ \\
\addlinespace
ET-2L & $410^{+39}_{-37}$ \\
\addlinespace
ET-$\Delta$$+$CE & $402^{+40}_{-36}$ \\
\addlinespace
ET-2L$+$CE & $387^{+32}_{-31}$ \\
\addlinespace
\bottomrule\bottomrule
\end{tabular*}

%% file: JSD_tabular.tex
\begin{tabular}{c @{}l@{} cccc c||c cccc}
\hline\hline
 & & \multicolumn{4}{c}{$m_1$ \textsc{Prior}} & & & \multicolumn{4}{c}{$m_2$ \textsc{Prior}} \\
\hline
 & & Default & Double Gaussian & Gaussian & Uniform & & & Default & Double Gaussian & Gaussian & Uniform \\
\hline\hline
\multirow{4}{*}{\rotatebox[origin=c]{90}{\parbox[c][8mm][c]{2cm}{\centering \textsc{Posterior}}}} & \multicolumn{1}{|l}{Default} & 0.68 & \textbf{0.20} & 0.57 & 0.77 & & & 0.81 & \textbf{0.04} & 0.22 & 0.58 \\
 & \multicolumn{1}{|l}{Double Gaussian} & 0.85 & \textbf{0.43} & 0.55 & 0.90 & & & 0.96 & \textbf{0.44} & 0.73 & 0.78 \\
 & \multicolumn{1}{|l}{Gaussian} & 0.81 & \textbf{0.35} & 0.54 & 0.87 & & & 0.93 & \textbf{0.26} & 0.59 & 0.73 \\
 & \multicolumn{1}{|l}{Uniform} & 0.69 & \textbf{0.21} & 0.57 & 0.78 & & & 0.83 & \textbf{0.04} & 0.25 & 0.60 \\
\hline
\end{tabular}

%% file: eos_parameters_table.tex
\begin{tabular}{l@{\hspace{1.5cm}}c@{\hspace{1.5cm}}c@{\hspace{1.5cm}}c}
\toprule\toprule
Dataset & $M_{\mathrm{TOV}}$ [$M_{\odot}$] & $R_{1.4}$ [km] & $p(3n_{\mathrm{sat}})$ [MeV fm$^{-3}$] \\
\midrule
Prior & $2.20^{+0.77}_{-0.82}$ & $13.41^{+1.99}_{-2.57}$ & $94^{+107}_{-77}$ \\
Heavy PSRs & $2.37^{+0.49}_{-0.34}$ & $13.74^{+1.47}_{-1.48}$ & $116^{+93}_{-58}$ \\
\addlinespace
\hline
\addlinespace
GW231109 (default) & $2.36^{+0.46}_{-0.34}$ & $13.35^{+1.70}_{-1.41}$ & $113^{+89}_{-57}$ \\
GW231109 (Gaussian) & $2.37^{+0.48}_{-0.38}$ & $13.51^{+1.53}_{-1.43}$ & $115^{+100}_{-55}$ \\
GW231109 (double Gaussian) & $2.35^{+0.44}_{-0.36}$ & $13.38^{+1.54}_{-1.49}$ & $112^{+86}_{-56}$ \\
GW231109 (uniform) & $2.36^{+0.46}_{-0.34}$ & $13.41^{+1.53}_{-1.37}$ & $113^{+89}_{-52}$ \\
GW231109 (QUR) & $2.32^{+0.40}_{-0.32}$ & $13.30^{+1.59}_{-1.56}$ & $108^{+74}_{-54}$ \\
GW231109 ($a_i \leq 0.4$) & $2.37^{+0.47}_{-0.36}$ & $13.58^{+1.47}_{-1.51}$ & $116^{+91}_{-57}$ \\
GW231109 (\texttt{XAS}) & $2.36^{+0.44}_{-0.34}$ & $13.39^{+1.64}_{-1.37}$ & $113^{+87}_{-55}$ \\
\addlinespace
\hline
\addlinespace
GW170817 & $2.26^{+0.36}_{-0.28}$ & $12.39^{+1.13}_{-1.34}$ & $90^{+68}_{-49}$ \\
GW190425 & $2.35^{+0.43}_{-0.34}$ & $13.42^{+1.46}_{-1.40}$ & $113^{+79}_{-55}$ \\
\addlinespace
\hline
\addlinespace
GW170817$+$GW190425 & $2.26^{+0.37}_{-0.25}$ & $12.38^{+1.11}_{-1.27}$ & $90^{+65}_{-47}$ \\
GW170817$+$GW231109 & $2.24^{+0.37}_{-0.26}$ & $12.36^{+1.21}_{-1.30}$ & $88^{+66}_{-51}$ \\
GW190425$+$GW231109 & $2.32^{+0.39}_{-0.31}$ & $13.15^{+1.38}_{-1.42}$ & $107^{+76}_{-48}$ \\
\addlinespace
\hline
\addlinespace
GW170817$+$GW190425$+$GW231109 & $2.24^{+0.37}_{-0.24}$ & $12.24^{+1.21}_{-1.20}$ & $87^{+62}_{-48}$ \\
\bottomrule\bottomrule
\end{tabular}